\documentclass[letterpaper]{article} % DO NOT CHANGE THIS
\newif\ifarxiv
\arxivtrue

\usepackage{aaai2026}  % DO NOT CHANGE THIS
\usepackage{times}  % DO NOT CHANGE THIS
\usepackage{helvet}  % DO NOT CHANGE THIS
\usepackage{courier}  % DO NOT CHANGE THIS
\usepackage[hyphens]{url}  % DO NOT CHANGE THIS
\usepackage{graphicx} % DO NOT CHANGE THIS
\usepackage{natbib}  % DO NOT CHANGE THIS AND DO NOT ADD ANY OPTIONS TO IT
\usepackage{caption} % DO NOT CHANGE THIS AND DO NOT ADD ANY OPTIONS TO IT
\usepackage{setspace}
\usepackage{algorithm}
\usepackage{algorithmic}
\usepackage{booktabs}
\usepackage{multirow}
\usepackage{amsmath}
\usepackage{tikz}
\usepackage{hyperref}
\usepackage{pgfplots}
\pgfplotsset{compat=1.18}

\usepackage{cleveref}

\crefname{figure}{Fig.}{Figs.}
\Crefname{figure}{Figure}{Figures}
\crefname{table}{Table}{Tables}
\crefname{equation}{Eq.}{Eqs.}
\crefname{section}{Sec.}{Secs.}
\usepackage{newfloat}
\usepackage{listings}
\DeclareCaptionStyle{ruled}{labelfont=normalfont,labelsep=colon,strut=off} % DO NOT CHANGE THIS
\floatstyle{ruled}
\newfloat{listing}{tb}{lst}{}
\floatname{listing}{Listing}
\title{EMG-UP: Unsupervised Personalization in Cross-User EMG Gesture Recognition}
\author{
    Nana Wang\textsuperscript{\rm 1}, 
    Suli Wang\textsuperscript{\rm 2},
    Gen Li\textsuperscript{\rm 1}, 
    Zhaoxin Fan\textsuperscript{\rm 1}, 
    }
\affiliations{
    \textsuperscript{\rm 1}School of Computer Science and Engineering, Beijing University of Aeronautics and Astronautics, China\\
    \textsuperscript{\rm 2}Department of Computer Science, Technische Universität Darmstadt, Germany
}

\begin{document}

\maketitle

\begin{abstract}
Cross-user electromyography (EMG)-based gesture recognition represents a fundamental challenge in achieving scalable and personalized human-machine interaction within real-world applications. Despite extensive efforts, existing methodologies struggle to generalize effectively across users due to the intrinsic biological variability of EMG signals, resulting from anatomical heterogeneity and diverse task execution styles. To address this limitation, we introduce EMG-UP, a novel and effective framework for Unsupervised Personalization in cross-user gesture recognition. The proposed framework leverages a two-stage adaptation strategy: (1) Sequence-Cross Perspective Contrastive Learning, designed to disentangle robust and user-specific feature representations by capturing intrinsic signal patterns invariant to inter-user variability, and (2) Pseudo-Label-Guided Fine-Tuning, which enables model refinement for individual users without necessitating access to source domain data. Extensive evaluations show that EMG-UP achieves state-of-the-art performance, outperforming prior methods by at least 2.0\% in accuracy.
\end{abstract}


\begin{figure}[!h]
  \centering
  \includegraphics[width=\columnwidth]{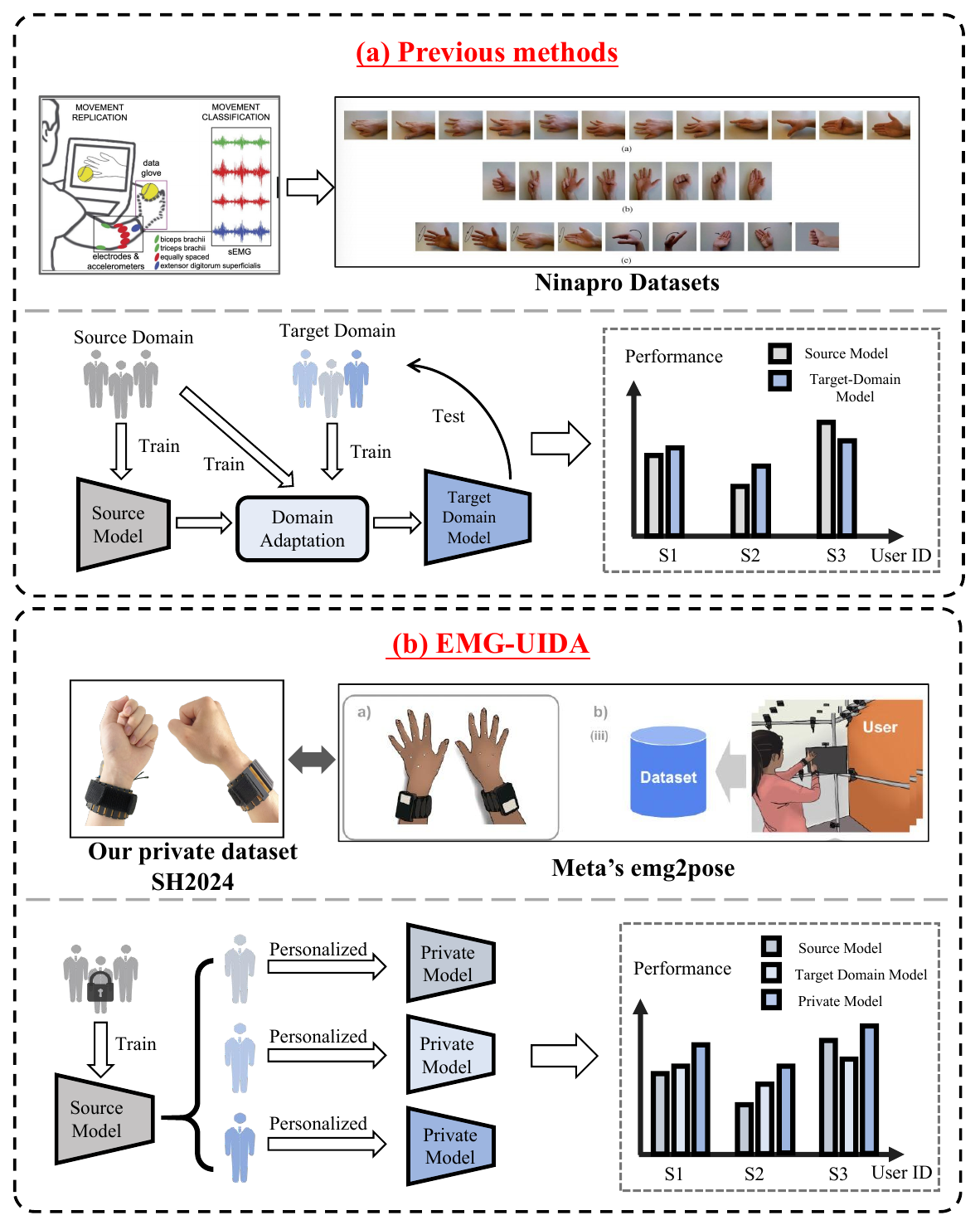}
  \caption{Conventional domain-adaptation pipelines vs. our EMG-UP personalization.  Previous methods rely on Domain Adaptation to obtain an adapted model, while our proposed approach focuses on personalization to derive a private, user-specific model. Both publicly available dataset and a private dataset is used to evaluate our framework.}
  \label{fig:teaser}
\end{figure}

\section{Introduction}
The ability to decode human gestures through electromyography (EMG) signals has redefined how we interact with machines, offering a direct pathway from muscle activity to actionable control. By interpreting the rich, high-dimensional electrical signals generated by human muscles, EMG-based gesture recognition systems unlock transformative possibilities in assistive robotics, prosthetic control, immersive interfaces, and rehabilitation \cite{tsinganos2018deep,lopez2024cnn,neaccsu2024emg,li2021gesture,hashi2024systematic,jiang2021emerging}. These systems not only promise intuitive and natural interaction but also pave the way for bridging the gap between biological signals and intelligent technology.

Traditional EMG gesture recognition methods~\cite{asif2020performance, jaramillo2020real, dhumal2021emg,ni2024survey} primarily focus on enhancing cross-class separability through feature engineering or advanced neural architectures. However, the biological variability of EMG signals remains a significant challenge, especially in cross-user scenarios. These variations stem from factors such as anatomical differences (e.g., muscle geometry, subcutaneous fat distribution) and diverse task execution styles~\cite{pradhan2022multi,mccrary2018emg}. Even with standardized electrode placement protocols, such variability is inherently unpredictable~\cite{ni2024survey,gowda2025database}, limiting the scalability and effectiveness of user-independent models when deployed across diverse populations.

Cross-user adaptation methods~\cite{wang2023iterative,zheng2022user,fu2024cross} attempt to address the challenge by calibrating models with background user data. However, these approaches suffer from critical limitations. First, they often assume that EMG signal distributions across users are homogeneous, which oversimplifies the complex and individualized nature of such data. This assumption leads to suboptimal adaptation, especially when signal discrepancies are substantial. Second, these methods typically require access to source domain data during the adaptation process, which raises privacy concerns and complicates deployment in real-world scenarios where source data may not be available. Moreover, existing adaptation strategies frequently rely on rigid, pre-defined pipelines that fail to generalize effectively to unseen users, further restricting their practical applicability. To overcome these limitations, our method abandons the assumption of signal homogeneity, removes the dependency on source data, and replaces fixed adaptation pipelines with a flexible, personalized approach that effectively generalizes to unseen users.

We propose \textbf{EMG-UP}, a novel framework for \textbf{Unsupervised Personalization} in cross-user EMG gesture recognition systems. Unlike traditional methods constrained by source domain dependence or simplistic adaptation strategies, EMG-UP introduces a principled, two-step adaptation pipeline designed to enable seamless personalization to new, unlabeled users. First, the framework employs Sequence-Cross Perspective Contrastive Learning to uncover robust, user-specific representations by capturing fine-grained variations across multiple perspectives. This step is followed by Pseudo-Label-Guided Fine-Tuning, which dynamically refines the model using the user's unique signal characteristics, ensuring optimal adaptation. By removing the need for source data access and prioritizing individualized customization, EMG-UP bridges the gap between model generalization and real-world deployment, offering a scalable solution for personalized, user-centric gesture recognition systems. To comprehensively evaluate our approach, we conduct extensive experiments on both publicly available EMG datasets and a proprietary dataset. The results demonstrate that EMG-UP achieves state-of-the-art performance, affirming its effectiveness and robustness in the real world. The key contributions of this work are summarized as follows:

\begin{itemize}
\item We propose {EMG-UP}, a novel framework for source-free unsupervised personalization in EMG-based gesture recognition. EMG-UP enables pre-trained models to adapt to new, unseen users without requiring access to source data, addressing the critical need for personalized adaptation and efficient deployment in the real world.

\item We introduce a {two-stage personalized adaptation strategy}, comprising Sequence-Cross Perspective Contrastive Learning and Pseudo-label-Guided Fine-Tuning. The first step captures inter-perspective variations to extract robust, user-specific features, while the second step refines the model using pseudo-labels, ensuring effective adaptation to individual users.  

\item We conduct extensive experiments on publicly available EMG datasets and our private dataset, demonstrating that EMG-UP achieves state-of-the-art performance.  

\end{itemize}
\section{Related Work}
\subsection{EMG-Based Gesture Recognition}

Early EMG-based gesture recognition relies on handcrafted features and classical classifiers, but recent progress is dominated by end-to-end deep architectures.

Convolutional-based methods capture local muscle activations, while recurrent or attention models capture long-range dependencies. ~\cite{pinzon2019convolutional} demonstrate that multichannel CNNs can automatically extract discriminative features from raw EMG data, while \cite{rahimian2020surface} extend this with a Hybrid Recognition Model combining CNNs and RNNs, as well as a Dilated Convolutional Network that reduces computational complexity. Temporal modeling also shows notable progress, with \cite{samadani2018gated} employing Bi-LSTM with attention mechanisms, and \cite{jabbari2020emg} developing adaptive LSTMs capable of generalizing across varying muscle contraction intensities. \cite{hu2018novel} further enhance this approach by incorporating attention mechanisms into CNN-RNN hybrids, enabling the automatic focus on discriminative signal segments and advancing the state of the art in EMG-based gesture recognition.  Multimodal fusion architectures~\cite{xie2018movement, zandigohar2024multimodal} further improves robustness by combining inertial or vision signals. 
% Recent advances in EMG-based gesture recognition emphasize the integration of signal processing techniques with deep learning architectures. 
% \citep{oh2021classification} combine Scale-Averaged Wavelet Transform (SAWT) with CNNs to enhance time-frequency features, while ~\cite{huang2019surface} introduce a CNN-LSTM model that effectively captures both spatial and temporal dynamics of EMG signals.  Another promising direction lies in multimodal fusion architectures. \cite{xie2018movement} and \cite{zandigohar2024multimodal} develop hierarchical C-RNN models that combine convolutional and recurrent layers, demonstrating the effectiveness of multimodal sensor fusion.  \cite{hu2018novel} further enhance this approach by incorporating attention mechanisms into CNN-RNN hybrids, enabling the automatic focus on discriminative signal segments and advancing the state of the art in EMG-based gesture recognition. 

Despite these advances, cross-user variability persists as a significant challenge, models trained on pooled data still degrade sharply when deployed to unseen users because EMG distributions are highly user-specific \cite{ni2024survey}. 
% To address this, our work proposes a robust framework for cross-user adaptation in EMG gesture recognition.

\subsection{Cross-user Adaptation}
% To bridge the user gap, three strategies dominate.
% \begin{itemize}
%     \item \textbf{Feature-level alignment} matches source/target statistics with spectral descriptors, adaptive sampling or MMD regularisation~\cite{hazarika2018automatic,zhang2022domain}.
%     \item \textbf{Model-level transfer} fine-tunes a pretrained network either with a few labels or via entropy-minimisation self-training~\cite{li2021transfer,hajian2024generalizing}.
%     \item \textbf{Adversarial or contrastive UDA} enforces user-invariant embeddings \cite{chen2021ms, raghu2025self}—e.g.\ OT-ST \cite{li2023cross} couples optimal transport with a student–teacher scheme, and STDA iteratively refines pseudo-labels plus discrepancy loss.  

% \end{itemize}

To bridge the user gap, cross-user adaptation seeks to address the variability in EMG signals caused by differences in anatomy, sensor placement, and muscle activation patterns among users. Traditional methods often focus on learning generalized models that work across users by leveraging techniques such as feature extraction \cite{hazarika2018automatic,zhang2022domain} and transfer learning \cite{bird2020cross,hajian2024generalizing}. These approaches aim to identify user-invariant representations or fine-tune pre-trained models on new users. In EMG-based gesture recognition, inter-session variability and differences in muscle activation dynamics present additional challenges, which have been tackled using signal processing techniques like time warping and filtering, as well as multi-view self-supervised learning (SSL) methods that leverage transformations such as channel masking to enhance robustness ~\cite{tsai2020self}. Pseudo-labeling methods, such as self-training with entropy minimization ~\cite{grandvalet2004semi}, have also been explored to enable unsupervised adaptation. For biomedical signals, recent work has integrated SSL with domain adaptation (DA) to improve cross-subject generalization \cite{raghu2025self,du2017semi,weng2024self}. For example,  \citep{chen2021ms} combined contrastive learning with Maximum Mean Discrepancy (MMD) to align cross-subject EEG data. While these methods demonstrate the potential of SSL and DA, they often rely on access to source data during adaptation, which raises concerns regarding privacy and practicality in real-world settings. Furthermore, most existing methods are designed for ECG or EEG data \cite{ding2024deep,wang2025robust} and lack explicit mechanisms to address EMG-specific challenges, such as variability across sessions and users \cite{du2017surface}.

% Unlike traditional approaches that require source data for domain adaptation, this paper focuses on unsupervised personalization, introducing a privacy-preserving framework to derive private, user-specific models. 

Although effective, these approaches need direct access to the source data during adaptation, raising privacy and deployment concerns in consumer applications.

\subsection{Source-Free Domain Adaptation}

Unlike classical Unsupervised Domain Adaptation, Source-Free DA assumes the original training data are no longer available and instead adapts a source-pretrained model to the target user only from unlabeled data~\cite{li2024comprehensive}. Early works such as SHOT~\cite{liang2020we} perform information-maximisation fine-tuning to align marginal feature distributions, while AdaContrast~\cite{chen2022contrastive} stabilises this step with contrastive prototypes. To further polish class boundaries, pseudo-label driven methods—such as NRC \cite{yang2021exploiting}, CPGA~\cite{qiu2021source}, iteratively filter noisy predictions. 
% Recent test-time approaches like TENT~\cite{wang2020tent} push SFDA toward real-time, on-device personalization.
In the biomedical domain, DFDA~\cite{liu2024novel} introduces dynamic feature alignment for cross-user EMG yet ignores label-space drift, and MS-MDA~\cite{chen2021ms} focuses on EEG emotion decoding.

Comprehensive surveys highlight the rapid expansion of SFDA to vision, speech and medical imaging~\cite{li2024comprehensive}. Yet, to our knowledge no study has explored source-free cross-user personalisation for sEMG. EMG-UP bridges this gap by combining a contrastive marginal-alignment stage with confidence-aware pseudo-labelling, thereby delivering privacy-preserving, per-user models without revisiting the source recordings.

% EMG-UP is the first to combine sequence-aware contrastive alignment with confidence-guided pseudo-label refinement for source-free, per-user sEMG gesture recognition, updating the entire backbone rather than merely calibrating a frozen head, thereby achieving robust personalization under strict privacy constraints.

\begin{figure*}[!h]
\centering
\includegraphics[width=\linewidth]{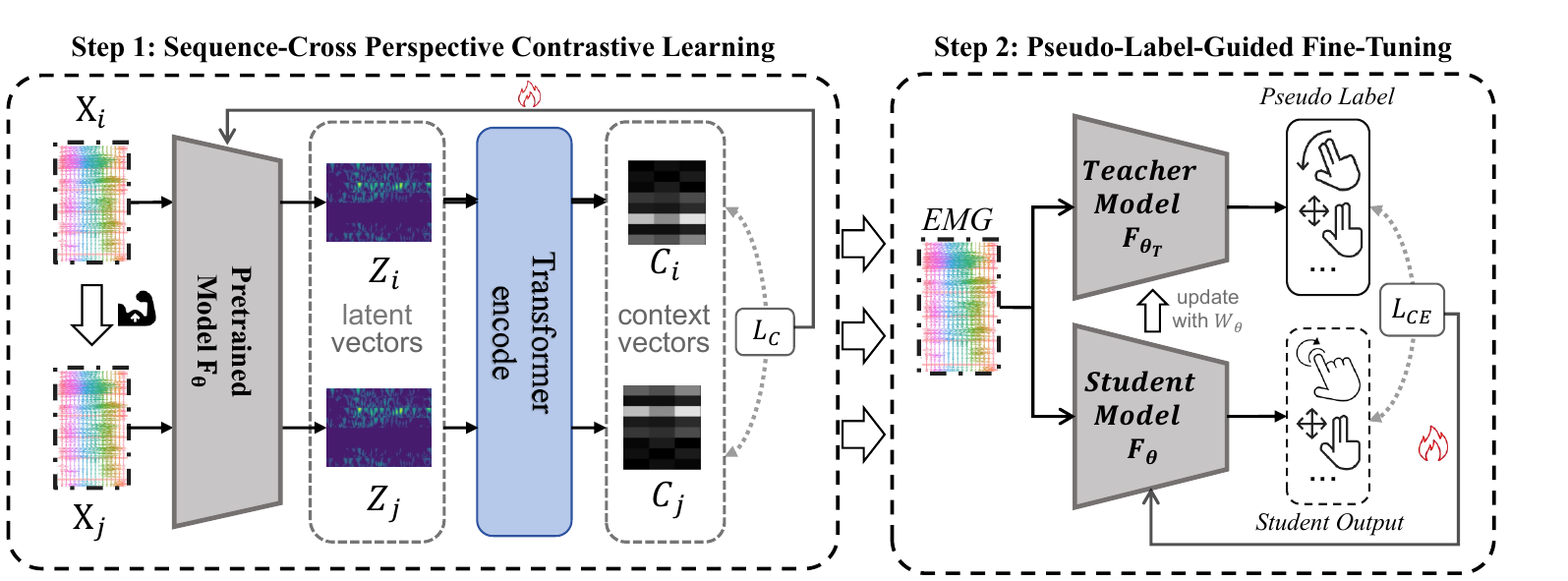}
\caption{The framework of our method. Our method consists of two stages: (1)Sequence-Cross Perspective Contrastive Learning, and (2)Pseudo-Label-Guided Fine-Tuning.}
\label{overview}
\end{figure*}

\section{Method}

\subsection{Problem Definition and Overview}
In real-world applications, the physiological state and gesture behavior of new users often differ significantly from the samples in the training data, which limits the model's generalization ability. When directly applied to new users, the model’s performance typically degrades. Traditional EMG-based gesture recognition models are usually trained and tested on data from the same dataset, failing to account for inter-individual differences. Therefore, it is crucial to develop models capable of adapting to new users under real-world conditions.

To address this challenge, we propose EMG-UP for EMG Gesture Recognition. Unlike traditional approaches that use joint training or domain adaptation methods requiring simultaneous access to both source and target data, EMG-UP personalizes the model for each new user using only their unlabeled data. By treating each user as an independent target domain, the framework introduces a flexible and plug-and-play adaptation mechanism that eliminates the need for source data. This is crucial for practical deployment scenarios where data privacy and user-specific adaptation are priorities. We define our task as follows:

Let the labeled source domain be defined as:
\begin{equation}
    \mathcal{D}_S = \{(X_S^i, Y_S^i)\}_{i=1}^{N_S},
\end{equation}
where \(X_S^i\) and \(Y_S^i\) represent the EMG signal and its corresponding label for the \(i\)-th sample in the source domain, and \(N_S\) is the number of labeled source samples.

The unlabeled target domain for a new user is represented as:
\begin{equation}
    \mathcal{D}_T = \{X_T^j\}_{j=1}^{N_T},
\end{equation}
where \(X_T^j\) denotes the EMG signal of the \(j\)-th sample in the target domain, and \(N_T\) is the number of target samples.

Our goal is to learn a personalized function \(F_T\) that maps the target domain data into a task-consistent feature space, without requiring access to source data or labels:
\begin{equation}
    F_T: X_T \rightarrow \mathcal{Z}, \quad \text{s.t.} \quad F_T \approx F_S \text{ in } \mathcal{Z}.
\end{equation}

Here, \(\mathcal{Z}\) denotes a latent feature space shared across users, and \(F_S\) is the pre-trained source model. This formulation reflects an unsupervised, cross-user adaptation setting where the goal is to align the target domain to the source domain's representation space without relying on source data at test time.

To solve the task, our proposed EMG-UP framework is designed to achieve user-specific adaptation in a source-free and unsupervised manner. The framework consists of two primary stages, as illustrated in Fig.~\ref{overview}. These stages are: (1) Sequence-Cross Perspective Contrastive Learning. This stage is designed to disentangle robust and user-specific feature representations. By capturing intrinsic EMG signal patterns invariant to inter-user variability, the model aligns the target domain's marginal distribution with the pre-trained source model. (2) Pseudo-Label-Guided Fine-Tuning. In this stage, high-confidence pseudo-labels are generated for the unlabeled target data. These pseudo-labels are used to refine the model further, enabling it to adapt to user-specific behaviors and align class-conditional distributions.

Through this two-stage process, EMG-UP enables efficient and privacy-preserving personalization, ensuring robust gesture recognition performance for unseen users. Next, we introduce the two stages in detail.

\subsection{Stage 1: Sequence-Cross Perspective Contrastive Learning}

The first stage of EMG-UP focuses on aligning the marginal distributions between the source and target domains while capturing user-specific feature representations. This is achieved through a sequence-cross-perspective contrastive learning strategy, which leverages augmented views of the target data.

Given an input EMG sequence \(X = (x_1, x_2, \dots, x_L)\), an augmented view \(X' = (x_L, x_{L-1}, \dots, x_1)\) is generated by reversing the sequence. Both \(X\) and \(X'\) are passed through the feature extractor and feature encoder to obtain their respective latent representations:
\begin{equation}
    Z = \{z_t\}_{t=1}^{L}, \quad Z' = \{z'_t\}_{t=1}^{L}.
\end{equation}

To capture temporal dependencies within the target domain, a Transformer-based autoregressive model is applied to encode context vectors \(C\) and \(C'\) for \(Z\) and \(Z'\):
\begin{equation}
    C = g(Z_{1:T}), \quad C' = g(Z'_{L-T:L}),
\end{equation}
where \(g(\cdot)\) represents the temporal encoder, and \(T\) is the context window size.

Subsequently, a sequence-cross prediction task is established, where \(C'\) is used to predict future time steps in \(Z\), and \(C\) is used to predict past time steps in \(Z'\). The learning objective is formulated as:
\begin{equation}
    L_C = -\frac{1}{K} \sum_{k=1}^{K} \log \frac{\exp(\text{sim}(\hat{z}_{T+k}, z_{T+k}))}{\sum_{z \in \mathcal{Z}} \exp(\text{sim}(\hat{z}_{T+k}, z))},
\end{equation}
where \(\text{sim}(\cdot, \cdot)\) denotes cosine similarity, \(\hat{z}_{T+k}\) is the predicted latent vector, and \(\mathcal{Z}\) is the set of all latent vectors.

\begin{figure}[!h]
  \centering
  \includegraphics[width=\linewidth]{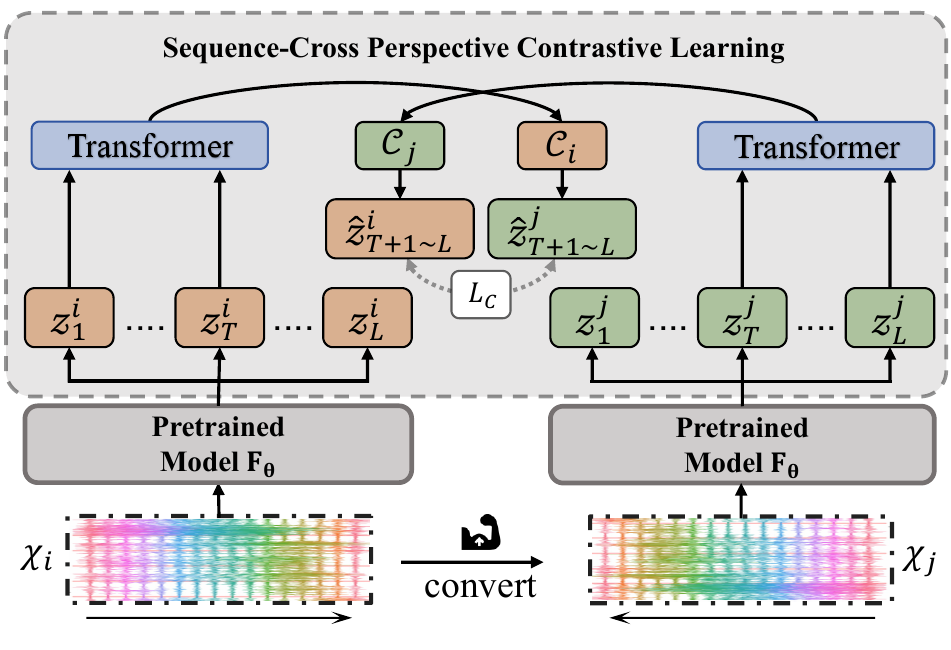}
   \vspace{-0.2in}
  \caption{Illustration of sequence-cross perspective contrastive learning. This module captures bidirectional relationships in EMG signal sequences through augmented views.}
  \label{fig:crossview}
  \vspace{-0.2in}
\end{figure}

This approach reduces the variability between the source and target domains while ensuring the model learns user-specific signal patterns, as shown in Fig.~\ref{fig:crossview}.

\subsection{Stage 2: Pseudo-Label-Guided Fine-Tuning}

Although the first stage aligns the marginal distributions, discrepancies in the class-conditional distributions (\(P_T(x|y)\)) between the source and target domains may still remain. To address this, the second stage employs a pseudo-label-guided fine-tuning strategy, which refines the model for individual users.

Using the pre-trained model as a teacher, pseudo-labels are generated for the target domain data:
\begin{equation}
    \hat{Y}_T = \text{argmax} \, F_\theta(X_T).
\end{equation}
To ensure pseudo-label quality, only predictions with confidence scores exceeding a predefined threshold \(\xi\) are retained:
\begin{equation}
    \text{Conf}(\hat{y}_i) > \xi,
\end{equation}
where \(\text{Conf}(\hat{y}_i)\) denotes the confidence score for the \(i\)-th pseudo-label.

The retained pseudo-labels are used to fine-tune the model with a cross-entropy loss:
\begin{equation}
    L_{CE} = -\frac{1}{N_T} \sum_{i=1}^{N_T} \hat{y}_i \log F_\theta(x_i).
\end{equation}

The teacher model is updated using an exponential moving average (EMA) of the student model parameters:
\begin{equation}
    \theta_T = \alpha \theta_T + (1 - \alpha) \theta,
\end{equation}
where \(\alpha\) is the EMA decay factor. This iterative process ensures that the model adapts effectively to user-specific behaviors, achieving robust personalization.

\begin{figure}[!h]
\centering
\includegraphics[width=\linewidth]{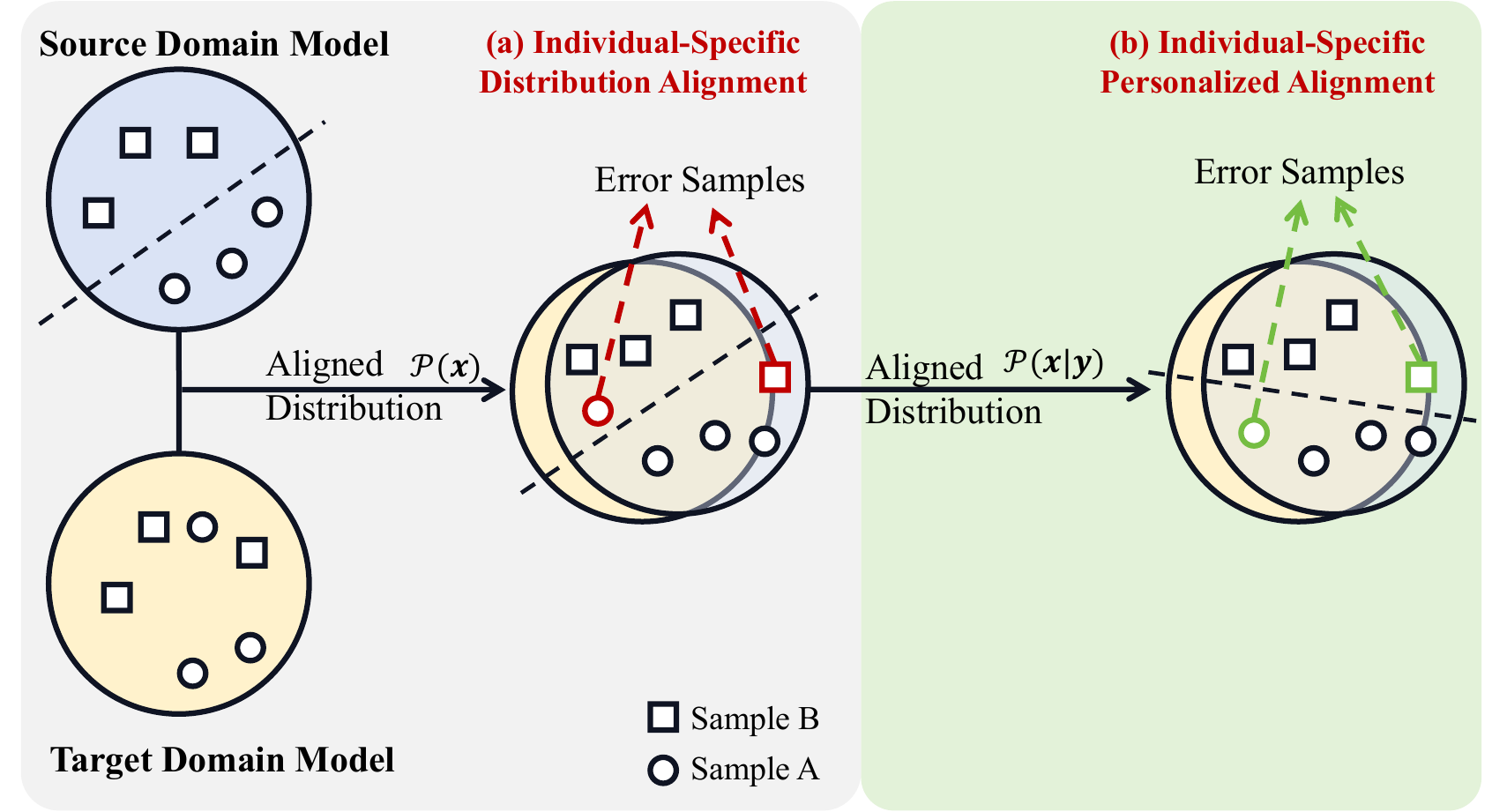}
\caption{Two Alignment Steps of EMG-UP Framework. This figure illustrates the two-stage aligment scheme of the proposed framework: (1) Individual-Specific
Distribution Alignment in Sequence-Cross Perspective
Contrastive Learning. (2) Individual-Specific
Personalized Alignment in Pseudo-Label-Guided Fine-Tuning.}
\label{fig:emg-up-framework}
\end{figure}

The above processes collectively constitute the Two-Stage Alignment Scheme of the EMG-UP Framework, as illustrated in Fig.~\ref{fig:emg-up-framework}. The complete process of EMG-UP is summarized in Algorithm~\ref{alg:EMG-UP}. This framework achieves cross-user EMG-based gesture recognition with unsupervised personalization through two core components: (1) Individual-Specific Distribution Alignment via Sequence-Cross Perspective Contrastive Learning, which disentangles robust and user-specific feature representations by capturing intrinsic signal patterns invariant to inter-user variability, and (2) Individual-Specific Personalized Alignment through Pseudo-Label-Guided Fine-Tuning, which refines the model to adapt to unique user-specific characteristics without requiring access to source domain data. Together, these two stages form a comprehensive alignment strategy that enables EMG-UP to simultaneously generalize across users and achieve high personalization.

\begin{algorithm}[h]
\caption{EMG-UP Framework}
\label{alg:EMG-UP}
\begin{algorithmic}[1]
\REQUIRE Labeled source domain data \(D_S = \{(X_S^i, Y_S^i)\}\), unlabeled target domain data \(D_T = \{X_T^j\}\)
\ENSURE Personalized target model \(F_\theta\)

\STATE \textbf{Step 1: Pre-train Source Model}
\STATE Train a source model \(F_\theta\) on \(D_S\) to learn general EMG signal features.

\STATE \textbf{Step 2: Sequence-Cross Perspective Contrastive Learning}
\FOR{each target sequence \(X_T^j \in D_T\)}
    \STATE Generate an augmented view \(X_T'^j\) by reversing \(X_T^j\).
    \STATE Extract latent representations \(Z_T^j\) and \(Z_T'^j\).
    \STATE Optimize \(F_\theta\) using the contrastive loss \(L_C\).
\ENDFOR

\STATE \textbf{Step 3: Pseudo-Label-Guided Fine-Tuning}
\STATE Initialize teacher model \(F_{\theta_T}\) with weights \(\theta_T \leftarrow \theta\).
\FOR{each target sequence \(X_T^j \in D_T\)}
    \STATE Generate pseudo-labels \(\hat{Y}_T^j\) for \(X_T^j\).
    \STATE Retain high-confidence predictions \(\text{Conf}(\hat{y}_i) > \xi\).
    \STATE Fine-tune \(F_\theta\) using \(L_{CE}\).
    \STATE Update teacher model parameters using EMA.
\ENDFOR
\STATE Return the personalized target model \(F_\theta\).
\end{algorithmic}
\end{algorithm}

\section{Experiment}

\begin{table*}[!ht]
\caption{Performance Comparison Between EMG-UP and Existing Baselines}
\label{tab:comparison}
\centering
% \small
\resizebox{\textwidth}{!}{
\begin{tabular}{lccccccccccc}
\toprule
\multirow{2}{*}{\textbf{Method}} & 
\multicolumn{4}{c}{\textbf{emg2pose Dataset}} & 
\multicolumn{4}{c}{\textbf{SH2024 Dataset}} & 
\multirow{2}{*}{\shortstack{Source\\Data}} & 
\multirow{2}{*}{\shortstack{Target\\Labels}} & 
\multirow{2}{*}{Notes} \\
\cmidrule(lr){2-5} \cmidrule(lr){6-9}
& \multicolumn{2}{c}{vemg2pose} & \multicolumn{2}{c}{neuropose} & \multicolumn{2}{c}{vemg2pose} & \multicolumn{2}{c}{neuropose} \\
\cmidrule(lr){2-3} \cmidrule(lr){4-5} \cmidrule(lr){6-7} \cmidrule(lr){8-9}
& ACC & MF1 & ACC & MF1 & ACC & MF1 & ACC & MF1 \\
\midrule
Source Only & 62.1 & 55.3 & 57.4 & 52.2 & 60.2 & 54.5 & 53.0 & 50.1 & $\surd$ & $\times$ & No adaptation \\
TL-AlexNet~(\citeyear{li2021transfer}) & 67.9 & 60.1 & 65.4 & 58.7 & 67.2 & 61.3 & 62.5 & 56.0 & $\surd$ & $\surd$ & Supervised transfer using MRD features \\
TSD~(\citeyear{khushaba2017framework}) & 65.3 & 57.3 & 61.7 & 55.5 & 63.6 & 58.2 & 57.4 & 53.3 & $\surd$ & $\times$ & Spectral domain adaptation \\
StyleTransfer~(\citeyear{anicet2020parkinson}) & 66.7 & 59.8 & 63.3 & 57.6 & 65.2 & 61.8 & 60.6 & 56.5 & $\surd$ & $\times$ & Feature-level style alignment \\
TL Pretrain~(\citeyear{tsinganos2021transfer}) & 68.6 & 60.9 & 65.6 & 58.9 & 66.5 & 62.4 & 61.2 & 57.2 & $\surd$ & $\times$ & Source-based unsupervised TL \\
DFDA~(\citeyear{liu2024novel}) & 69.2 & 62.0 & 66.7 & 60.3 & 68.1 & 63.5 & 63.4 & 59.0 & $\surd$ & $\surd$ & UDA with dynamic feature alignment\\
DANN~(\citeyear{lee2024emg}) & 65.9 & 58.5 & 63.9 & 56.5 & 64.4 & 57.0 & 62.9 & 55.5 & $\surd$ & $\times$ & Classic adversarial UDA \\
SCDEM~(\citeyear{wang2024optimization}) & 67.6 & 60.0 & 66.4 & 58.8 & 66.6 & 59.0 & 65.4 & 57.8 & $\surd$ & $\times$ & The current SOTA UDA method \\
\textbf{EMG-UP (Ours)} & \textbf{71.4} & \textbf{64.3} & \textbf{69.2} & \textbf{63.8} & \textbf{70.1} & \textbf{65.5} & \textbf{64.5} & \textbf{60.2} & $\times$ & $\times$ & Source-free, unsupervised personalization \\
\bottomrule
\end{tabular}
}

\end{table*}

\subsection{Implementation Details}

\noindent \textbf{Datasets:} Although the Ninapro dataset is one of the most widely used benchmarks in EMG research, we do not include it in our evaluation for several reasons. First, most versions of Ninapro are collected with clinical-grade equipment under constrained lab settings, which limits their generalizability to real-world, consumer-grade applications. Second, the dataset lacks sufficient inter-session and cross-user variability required to evaluate personalized adaptation methods like EMG-UP. Clinical-grade systems~\cite{amma2015advancing, du2017surface} offer high fidelity but are impractical for scalable use due to their complex setup. Consumer-grade datasets, such as those collected using Myo~\cite{palermo2017repeatability, lobov2018latent}, are more accessible but limited by low resolution (e.g., 8 channels at 200 Hz), which restricts their utility for fine-grained modeling. These limitations motivate the choice of the \textbf{emg2pose} dataset~\cite{salter2024emg2pose} and our private dataset \textbf{SH2024}, as both address these shortcomings effectively. The emg2pose dataset includes 193 users and 751 sessions, recorded with a 16-channel wearable device at over 2 kHz, providing extensive variability across users and sessions. This makes it ideal for studying unsupervised cross-user personalization. The SH2024 dataset, on the other hand, contains 500 EMG sequences from 27 users across multiple sessions, covering 13 gesture classes. It complements emg2pose by enabling controlled validation.

\noindent \textbf{Evaluation Metrics:} We employ 10-fold cross-validation (CV) to assess the performance of our method on three different datasets. Following prior work \cite{lee2023stretchable}, we partition each dataset into a training set, validation set, and test set with a ratio of 8:1:1. The test set consists of unseen individuals, ensuring that there are no overlapping subjects between the training and test sets. Each individual appears in the test set only once across the 10-fold CV experiments. In each fold, we pre-train the source model using the training and validation sets. Subsequently, the source model is personalized for each individual in the test set. Finally, we compute the average evaluation metrics for all individuals in the test set. The evaluation metrics include accuracy (ACC) and macro F1-score (MF1).

\noindent \textbf{Implementation Details:}
All experiments, including the proposed EMG-UP method and all baseline methods, are implemented using PyTorch and trained on an 8-GPU GeForce RTX 3090 Ti setup.

All gesture classification tasks are performed using a pre-trained Convolutional Neural Network (CNN) as the backbone recognition model. The CNN is trained on the source domain data and remains fixed across all methods. Our proposed EMG-UP framework operates as a plug-in personalization module on top of this pre-trained CNN, enabling cross-user adaptation without accessing source data. This design ensures a fair comparison across all baseline and transfer learning methods, isolating the effect of the personalization strategy from model architecture variations.
 
For EMG-UP, the pre-training phase runs for 100 epochs with a learning rate of \( 1\mathrm{e}{-4} \). In the second stage, fine-tuning is conducted for 5 epochs with a learning rate of \( 1\mathrm{e}{-7} \) and a time step \( T \) set to 17. In the third stage, fine-tuning lasts for 10 epochs with a learning rate of \( 1\mathrm{e}{-7} \) and a momentum coefficient \( \alpha \) of 0.996. Additionally, we use the Adam optimizer with \( \beta \) parameters set to [0.5, 0.99], weight decay set to \( 3\mathrm{e}{-4} \), and a batch size of 64.

\subsection{Comparison Results}

We evaluate the performance of EMG-UP against a range of state-of-the-art methods for cross-user EMG-based gesture recognition, including TL-AlexNet~\cite{li2021transfer}, TSD~\cite{khushaba2017framework}, StyleTransfer~\cite{anicet2020parkinson}, TL Pretrain~\cite{tsinganos2021transfer}, DFDA~\cite{liu2024novel}, DANN~\cite{lee2024emg}, and SCDEM~\cite{wang2024optimization}. The "Source Only" baseline is also included, where no domain adaptation or personalization is applied. To ensure a fair comparison, all methods use the same backbone architecture and training settings.

As shown in Table 1, EMG-UP consistently outperforms all baselines across both datasets and evaluation metrics. On the emg2pose dataset, EMG-UP achieves 71.4\% accuracy and 64.3\% macro-F1 on vemg2pose, and 69.2\% and 63.8\% on neuropose, surpassing Source Only by more than 9\% on average. On SH2024, EMG-UP also leads with 70.1\% and 65.5\% (vemg2pose), and 64.5\% and 60.2\% (neuropose), showing robust generalization across both datasets.

Compared with the strongest UDA baselines such as DFDA and SCDEM, EMG-UP maintains a 2--3\% margin in both ACC and MF1. Unlike these methods, EMG-UP does not rely on access to source data or target labels, highlighting its practical advantage in privacy-sensitive or personalized deployment scenarios.

These results validate the effectiveness of our source-free, unsupervised personalization approach and demonstrate EMG-UP's strong potential for scalable real-world EMG-based interaction.EMG-UP’s strong performance stems from its \textbf{two-stage unsupervised personalization framework}, which tackles inter-user variability, source data absence, and poor generalization. In Stage I, \textbf{Sequence-Cross Perspective Contrastive Learning} extracts user-invariant features while preserving individual differences, aligning the target domain with the source model without supervision. In Stage II, \textbf{Pseudo-Label-Guided Fine-Tuning} enables user-specific adaptation through high-confidence pseudo labels, without requiring ground-truth annotations. By eliminating rigid calibration and source data dependency, EMG-UP generalizes well to new users and datasets, as confirmed by its consistent results on \textbf{emg2pose} and \textbf{SH2024}.

% \begin{table}[h]
%     \caption{Ablation Study Results of EMG-UP}
%     \label{tab:ablation}
%     \centering
%     \small
%     \setlength{\tabcolsep}{2.0pt} % Reduce horizontal padding
%     \renewcommand{\arraystretch}{1.1} % Reduce row height
%     \begin{tabular}{lcccccccc}
%         \toprule
%         \multirow{2}{*}{\textbf{Method}} & \multicolumn{4}{c}{\textbf{emg2pose Dataset}} & \multicolumn{4}{c}{\textbf{SH2024 Dataset}} \\
%         \cmidrule(lr){2-5} \cmidrule(lr){6-9}
%         & \multicolumn{2}{c}{\textit{vemg2pose}} & \multicolumn{2}{c}{\textit{neuropose}} & \multicolumn{2}{c}{\textit{vemg2pose}} & \multicolumn{2}{c}{\textit{neuropose}} \\
%         & ACC & MF1 & ACC & MF1 & ACC & MF1 & ACC & MF1 \\
%         \midrule
%         SO            & 62.1 & 55.3 & 57.4 & 52.2 & 60.2 & 54.5 & 53.0 & 50.1 \\
%         SO + SSA      & 69.6 & 63.1 & 67.7 & 61.2 & 69.8 & 64.5 & 63.4 & 58.3 \\
%         SO + SSP      & 69.2 & 62.8 & 68.4 & 61.6 & 69.4 & 64.8 & 63.7 & 58.9 \\
%         SO + SSA + SSP & \textbf{71.4} & \textbf{64.3} & \textbf{69.2} & \textbf{63.8} & \textbf{70.1} & \textbf{65.5} & \textbf{64.5} & \textbf{60.2} \\
%         \bottomrule
%     \end{tabular}
% \end{table}

\subsection{Ablation Study}

To further validate the advantages of the EMG-UP method, we conduct a series of ablation experiments to investigate the impact of different alignment strategies on model performance. The ablation study aims to evaluate how the model performs when specific components are removed and to confirm the contribution of each module to the overall framework. Additionally, we pay particular attention to the importance of the two-step alignment strategy (SSA and SSP) and analyze its effect on unsupervised individual domain adaptation through experiments.

To systematically evaluate the role of different alignment strategies, we define the following four model variants:
\begin{itemize}
    \item \textbf{SO (Source Only)}: The model is tested using only the source domain without any alignment strategy.
    \item \textbf{SO + SSA (Source Only + Subject-Specific Alignment)}: Retains only the individual distribution alignment (SSA) in the EMG-UP framework, which aligns the marginal distribution between the source and target domains.
    \item \textbf{SO + SSP (Source Only + Subject-Specific Personalization)}: Retains only the subject-specific personalization (SSP) in the EMG-UP framework, aligning the class-conditional distribution in the target domain.
    \item \textbf{SO + SSA + SSP (Full EMG-UP Framework)}: Utilizes the full two-step alignment strategy in the EMG-UP framework, combining both SSA and SSP modules.
\end{itemize}

\begin{table}[h]
    \caption{Ablation Study of EMG-UP. SO represents Source Only. SSA represents Subject-Specific Alignment. SSP represents Subject-Specific Personalization.}
    \label{tab:ablation}
    \centering
    \small
    \setlength{\tabcolsep}{2.5pt} % Reduce horizontal padding
    \renewcommand{\arraystretch}{1.1} % Reduce row height
    \begin{tabular}{lllcccccccc}
        \toprule
        \multicolumn{3}{c}{\textbf{Modules}} & \multicolumn{4}{c}{\textbf{emg2pose Dataset}} & \multicolumn{4}{c}{\textbf{SH2024 Dataset}} \\
        \cmidrule(lr){1-3} \cmidrule(lr){4-7} \cmidrule(lr){8-11}
        \multirow{2}{*}{SO} & \multirow{2}{*}{SSA} & \multirow{2}{*}{SSP} & \multicolumn{2}{c}{\textit{vemg2pose}} & \multicolumn{2}{c}{\textit{neuropose}} & \multicolumn{2}{c}{\textit{vemg2pose}} & \multicolumn{2}{c}{\textit{neuropose}} \\
        & & & ACC & MF1 & ACC & MF1 & ACC & MF1 & ACC & MF1 \\
        \midrule
        $\surd$ &        &        & 62.1 & 55.3 & 57.4 & 52.2 & 60.2 & 54.5 & 53.0 & 50.1 \\
        $\surd$ & $\surd$ &        & 69.6 & 63.1 & 67.7 & 61.2 & 69.8 & 64.5 & 63.4 & 58.3 \\
        $\surd$ &        & $\surd$ & 69.2 & 62.8 & 68.4 & 61.6 & 69.4 & 64.8 & 63.7 & 58.9 \\
        \textbf{$\surd$} & \textbf{$\surd$} & \textbf{$\surd$} & \textbf{71.4} & \textbf{64.3} & \textbf{69.2} & \textbf{63.8} & \textbf{70.1} & \textbf{65.5} & \textbf{64.5} & \textbf{60.2} \\
        \bottomrule
    \end{tabular}
\end{table}

% \begin{figure*}[!t]
%     \centering
%     \includegraphics[width=\linewidth]{pic/emg_uida_fig2_tsne_final.pdf}
%     \caption{
%         Visualization of target domain feature distributions before and after applying SSA. 
%         Without SSA (left), features from different gesture classes are loosely clustered and often overlap. 
%         After SSA (right), features are more compact and better separated across classes, 
%         demonstrating the effectiveness of cross-user distribution alignment. This aligns with the performance improvements shown in Table~\ref{tab:ablation}.
%     }
%     \label{fig:tsne_ssa}
% \end{figure*}

As shown in Table 2, the ablation results confirm the effectiveness of EMG-UP’s two-step alignment strategy for unsupervised cross-user personalization. Both the SSA and SSP modules significantly improve performance over the Source Only baseline, demonstrating their role in capturing individual-specific features. The SSP module slightly outperforms SSA, suggesting that class-conditional alignment better captures user gesture patterns. The full EMG-UP model, combining both modules, achieves the best results, highlighting their complementary benefits in handling cross-user variability and enhancing personalization.

% Fig.~\ref{fig:tsne_ssa} further illustrates the target domain feature distributions before (left) and after (right) applying SSA. Without SSA, features from different gesture classes are loosely clustered and exhibit significant overlap. After applying SSA, the features become more compact and distinctly separated, demonstrating the effectiveness of cross-user distribution alignment in improving class discrimination. This observation aligns with the performance improvements reported in Table~\ref{tab:ablation}.

% \begin{figure}[!ht]
%     \centering
%     \begin{tikzpicture}
%         \begin{axis}[
%             ybar,
%             bar width=10pt,
%             enlargelimits=0.15,
%             ylabel={Performance (\%)},
%             symbolic x coords={SO, SO+SSA, SO+SSP, Full},
%             xtick=data,
%             nodes near coords,
%             legend style={at={(0.5,-0.15)}, anchor=north, legend columns=-1},
%             ymin=50, ymax=75
%         ]
%         \addplot coordinates {(SO,60.2) (SO+SSA,69.8) (SO+SSP,69.4) (Full,70.1)};
%         \addplot coordinates {(SO,54.5) (SO+SSA,64.5) (SO+SSP,64.8) (Full,65.5)};
%         \legend{ACC, MF1}
%         \end{axis}
%     \end{tikzpicture}
%     \caption{Ablation study results (SH2024 dataset): ACC and MF1 under different alignment configurations.}
%     \label{fig:ablation_bar}
% \end{figure}

\paragraph{Impact of Confidence Filtering in SSP}

To further analyze the effectiveness of the pseudo-label selection strategy in the SSP module, we conduct an ablation experiment by removing the confidence-based filtering mechanism. Specifically, instead of discarding low-confidence sequences, all generated pseudo-labels are retained for fine-tuning. As shown in Table 3, model performance drops notably without confidence filtering, confirming that noisy pseudo-labels can degrade adaptation quality. This highlights the importance of the sequence-level confidence strategy ($\xi=0.8$, $N_c=15$) for stable and reliable personalization.

\begin{table}[!h]
\caption{Impact of Confidence Filtering in SSP.}
\centering
\small

\label{tab:confidence_filtering}
\begin{tabular}{lcc}
\hline
\textbf{Method} & \textbf{ACC} & \textbf{MF1} \\
\hline
SO + SSP (with filtering) & \textbf{69.4} & \textbf{64.8} \\
SO + SSP (no filtering)   & 66.2 & 61.5 \\
\hline
\end{tabular}

\end{table}

\vspace{2mm}
\paragraph{Sensitivity to Time Step $T$}

We also investigate the sensitivity of the SSA module to the hyperparameter $T$, which defines the temporal cutoff for constructing context windows in sequential cross-view contrastive learning. Table 4 presents model performance across various $T$ values. We observe that both too short and too long time windows lead to suboptimal results, while $T=17$ offers the best trade-off between context richness and temporal consistency.

\begin{table}[!h]
\caption{Sensitivity Analysis on Time Step $T$ (SO + SSA setting. )}
\centering
\small
\label{tab:time_step_sensitivity}
\begin{tabular}{ccc}
\hline
\textbf{Time Step $T$} & \textbf{ACC} & \textbf{MF1} \\
\hline
8  & 67.5 & 61.4 \\
12 & 68.2 & 62.9 \\
\textbf{17} & \textbf{69.6} & \textbf{64.5} \\
24 & 68.1 & 62.1 \\
\hline
\end{tabular}

\end{table}

\vspace{2mm}
\paragraph{Effect of Sequence Inversion Strategy}

Our SSA module leverages sequence inversion to create cross-view pairs for contrastive learning. To evaluate the impact of this design, we compare it against two variants: (1) \textit{SSA w/o inversion}, which uses only the original sequence; and (2) \textit{SSA w/ random masking}, which replaces inversion with time-series dropout masking. As shown in Table~\ref{tab:inversion_ablation}, the inversion-based view generation consistently yields superior results, demonstrating its effectiveness in capturing bidirectional temporal structure in EMG signals.

\begin{table}[h]
\caption{Effect of Sequence Inversion in SSA.}
\centering
\small
\label{tab:inversion_ablation}
\begin{tabular}{lcc}
\hline
\textbf{Method} & \textbf{ACC} & \textbf{MF1} \\
\hline
SSA w/ inversion (ours) & \textbf{69.6} & \textbf{64.5} \\
SSA w/o inversion       & 66.8 & 61.2 \\
SSA w/ random masking   & 67.5 & 62.1 \\
\hline
\end{tabular}

\vspace{-0.2in}
\end{table}

\section{Conclusion}

% Cross-user EMG-based gesture recognition remains a critical challenge in achieving scalable and personalized human-machine interaction for real-world applications. The intrinsic biological variability of EMG signals, stemming from anatomical heterogeneity and diverse task execution styles, significantly hinders the generalization capability of existing methods. To address these challenges, we proposed EMG-UP, a novel framework for Unsupervised Personalization in cross-user gesture recognition. The EMG-UP framework introduces a two-stage adaptation strategy: (1) Sequence-Cross Perspective Contrastive Learning, which disentangles robust and user-specific feature representations by capturing intrinsic signal patterns invariant to inter-user variability, and (2) Pseudo-Label-Guided Fine-Tuning , which refines model performance for individual users without requiring access to source domain data. These complementary strategies enable EMG-UP to effectively align inter-user distributions while maintaining high personalization capabilities. Comprehensive evaluations on both widely adopted public benchmarks and a proprietary EMG dataset demonstrate that EMG-UP achieves state-of-the-art performance. The results validate the framework's robustness, scalability, and effectiveness in addressing cross-user variability, paving the way for more reliable and personalized EMG-based human-machine interaction systems in real-world scenarios.

This paper introduces EMG-UP, a source-free framework for unsupervised personalization in cross-user EMG-based gesture recognition. EMG-UP unites sequence-aware contrastive alignment with confidence-guided pseudo-label refinement, updating the entire backbone rather than merely fine-tuning a frozen classification head. This holistic adaptation enables robust per-user performance while fully respecting data-privacy constraints.

The framework proceeds in two stages: (1) {Sequence-Cross Perspective Contrastive Learning} isolates representations that are simultaneously robust and user-specific by capturing intrinsic signal patterns that remain invariant across users. (2) {Pseudo-Label-Guided Fine-Tuning} then sharpens class boundaries for each individual without ever revisiting the source domain.

EMG-UP consistently outperforms all baselines across both datasets and evaluation metrics. On the \textbf{emg2pose} dataset, EMG-UP achieves \textbf{71.4\%} accuracy and \textbf{64.3\%} macro-F1 on the \textit{vemg2pose} representation, and \textbf{69.2\%} accuracy and \textbf{63.8\%} macro-F1 on \textit{neuropose}, surpassing the Source Only baseline by over \textbf{9\% on average}. Similarly, on the \textbf{SH2024} dataset, EMG-UP attains \textbf{70.1\%} / \textbf{65.5\%} (vemg2pose) and \textbf{64.5\%} / \textbf{60.2\%} (neuropose) in accuracy and macro-F1 respectively, demonstrating that EMG-UP achieves state-of-the-art performance in terms of classification accuracy and personalization robustness, outperforming conventional transfer learning and domain adaptation methods.

Looking forward, we aim to extend EMG-UP to real-time, in-the-wild environments, where EMG signals exhibit higher variability due to factors such as motion artifacts, long-term drift, and unpredictable usage conditions. This will involve developing lightweight inference modules, continual learning pipelines, and online calibration techniques to support adaptive and seamless interaction in practical applications such as robotics dexterous hand control, AR/VR interfaces, and gaming systems. By bridging the gap between algorithmic innovation and deployment feasibility, EMG-UP paves the way for scalable and privacy-preserving EMG interfaces in next-generation human-computer interaction systems.

\bibliography{aaai2026}
% \end{spacing}
\makeatletter
\@ifundefined{isChecklistMainFile}{
  % We are compiling a standalone document
  \newif\ifreproStandalone
  \reproStandalonetrue
}{
  % We are being \input into the main paper
  \newif\ifreproStandalone
  \reproStandalonefalse
}
\makeatother

\ifreproStandalone
\documentclass[letterpaper]{article}
\usepackage[submission]{aaai2026}
\setlength{\pdfpagewidth}{8.5in}
\setlength{\pdfpageheight}{11in}
\usepackage{times}
\usepackage{helvet}
\usepackage{courier}
\usepackage{xcolor}
\frenchspacing

\begin{document}
\fi
\setlength{\leftmargini}{20pt}
\makeatletter\def\@listi{\leftmargin\leftmargini \topsep .5em \parsep .5em \itemsep .5em}
\def\@listii{\leftmargin\leftmarginii \labelwidth\leftmarginii \advance\labelwidth-\labelsep \topsep .4em \parsep .4em \itemsep .4em}
\def\@listiii{\leftmargin\leftmarginiii \labelwidth\leftmarginiii \advance\labelwidth-\labelsep \topsep .4em \parsep .4em \itemsep .4em}\makeatother

\setcounter{secnumdepth}{0}
\renewcommand\thesubsection{\arabic{subsection}}
\renewcommand\labelenumi{\thesubsection.\arabic{enumi}}

\newcounter{checksubsection}
\newcounter{checkitem}[checksubsection]

\newcommand{\checksubsection}[1]{%
  \refstepcounter{checksubsection}%
  \paragraph{\arabic{checksubsection}. #1}%
  \setcounter{checkitem}{0}%
}

\newcommand{\checkitem}{%
  \refstepcounter{checkitem}%
  \item[\arabic{checksubsection}.\arabic{checkitem}.]%
}
\newcommand{\question}[2]{\normalcolor\checkitem #1 #2 \color{blue}}
\newcommand{\ifyespoints}[1]{\makebox[0pt][l]{\hspace{-15pt}\normalcolor #1}}

\end{document}
\fi
\end{document}